\documentclass[onecolumn,showpacs,preprintnumbers,amsmath,amssymb]{revtex4}

\usepackage{graphicx}
\usepackage{dcolumn}
\usepackage{bm}

\begin{document}

\title{Dirac Equation in External Non-Abelian Gauge Field in Plane Wave Approximation }

\author{A.V.Koshelkin.}
 \altaffiliation[  koshelkin@mtu-net.ru; koshelkin@theor.mephi.ru; ]
 {}
\affiliation{Moscow Institute for Physics and Engineering,
Kashirskoye sh., 31, 115409 Moscow, Russia }
\date{\today}

\begin{abstract}
The exact solutions of the Dirac equation in an external
non-abelian $SU(N)$ gauge field which is in the form of a plane
wave  on the light cone is obtained. The whole set of the
solutions for both particles  and anti-particle is constructed.
\end{abstract}


\maketitle

\section{Introduction}

The study of  non-abelian gauge fields plays an important  role in
the modern field theory\cite{1,2,3}.  This primarily concerns the
field governed by unitary symmetries. Exactly such symmetry
($SU(N)$) is the basis of QCD\cite{4} where quarks form the space
of the fundamental representation of $SU_c(3)$ group while  gauge
fields provide interaction between them.  Therefore the knowledge
of the solution of the Dirac equations in such field is the key to
understanding the processes taking place in the strong interacting
matter appearing, particular,  in the collisions of heavy ions of
high energies\cite{kh}.

The Dirac equation in the field  of a plane electromagnetic field
wave  has been studied  by D.M.Volkov who has  derived  the exact
solution\cite{6} for the first time. The key point in obtaining
the such solution is the commutativity of electromagnetic field.

In the present paper the exact solution of the Dirac equation in
an external non-abelian $SU(N)$ gauge field  which  represents  a
plane wave on the light cone is obtained. The whole set of the
solutions which determines the states of  both particles and
anti-particles is derived. On a basis of the obtained  solutions
the fermion current is calculated. It strongly  differs in
structure from the fermion current in the field of a plane wave in
QED\cite{6} .

The paper is organized as follows. The statement of the considered
problem is formulated in Section II. Section III contains the
solution of the Dirac equation in an external non-abelian gauge
field. The last section is conclusion. The transformation of the
operator exponent is presented in Appendix.

\section{Statement of the problem}

We  consider a fermion  with  a spin  $s=1/2$ in an external
non-abelian field $A_a^{\nu}$ . The fermion  fields $\Psi (x)$ and
 field $A_a^{\nu}$ form the space  of the fundamental ($\Psi
(x)$) and associated ($A_a^{\nu}$) representations of $SU(N)$
group, respectively. Then the fermion field $\Psi (x)$ satisfies
the equations\cite{7,8}:

\begin{widetext}
\begin{eqnarray}
&& \left\{ i \gamma^{\mu} \left( \partial_{\mu} + i g \cdot
A_{\mu}^a (x) T_a \right) - m \right\} \Psi (x) = 0  \\ \nonumber  \\
&& {\bar \Psi} (x) \left\{ i \gamma^{\mu} \left(
{\overleftarrow\partial}_{\mu} - i g \cdot {A}_{\mu}^a (x) T_a
\right) + m \right\}  = 0; \ \ \ \ \ {\bar \Psi} (x) = \Psi^\dag (
x ) \gamma^0 ,
\end{eqnarray}
\end{widetext}
where $m$ is a fermion mass, $g$ is the coupling constant;
$\gamma^{\nu}$ are the Dirac matrixes, $x^\mu = (x^0 ; {\vec x})$
is a vector in the Minkowski space-time; $\partial_{\mu} =
(\partial /\partial t ; \nabla )$; the Roman letters numerate the
basis in the space of the associated representation of $SU(N)$
group, so that $a,b,c = 1 \dots N^2 -1$. We use the signature $
diag \left( {\cal G}^{\mu \nu} \right) = (1; -1; -1; -1)$ for the
metric tensor ${\cal G }^{\mu \nu}$. The line over $\Psi$ and
"dagger"  mean the  Dirac and hermitian  conjugation,
respectively\cite{6,7}. Summing over any pair of repeated indexes
is implied.

The symbols $T_a$ in Eqs.(1), (2) are the generators of $SU(N)$
group which satisfy the commutative relations and normalization
condition:

\begin{widetext}
\begin{eqnarray}
&& \left[ T_a, T_b \right]_-    =  T_a  T_b - T_b T_a =   i f_{ab}^{\ \ c} T_c ;   \\
&& Tr \ ( T_a \ T_b )   ={1\over 2} \delta_{ab} ;
\end{eqnarray}
\end{widetext}
where $f_{ab}^{\ \ c}$ are the structure constant of the $SU(N)$
group, which are real and anti-symmetrical with respect to the
transposition in any pair of indexes;   $\delta_{ab}$ is the
Kroneker symbol. In the matrix representation the operators $T_a$
coincide with the Pauli and Gill-Mann matrixes when $N$ is equal
to $2$ or $3$, respectively

It directly follows from Eqs.(3), (4) that

\begin{widetext}
\begin{eqnarray}
&& \left[ T_a, T_b \right]_+ = T_a  T_b + T_b T_a =  {N^2 -2 \over
2 N} \delta_{ab} . \end{eqnarray}
\end{widetext}

Let the  fields $A_{\mu}^a (x)$ be dependent  on the 4-coordinate
$x$ in the form of a plane wave:

\begin{widetext}
\begin{eqnarray}
&& A_{\mu}^a (x) =  A_{\mu}^a  ( k x ) \equiv A_{\mu}^a  (\varphi
) ; \ \ \ \  k^\mu k_\mu = 0 ; \ \ \ \ \varphi = k_\mu x^\mu
\equiv k x ,
\end{eqnarray}
\end{widetext}
where $k=(k^0 ; {\vec k})$ is some 4-vector lying on the light
cone.

We fix the axial gauge by the expressions:

\begin{widetext}
\begin{eqnarray}
&& \partial^{\mu} A_{\mu}^a  = 0 ;  \ \ \ \ \ \ \ \ k^{\mu} {\dot
A}_{\mu}^a  = 0,
\end{eqnarray}
\end{widetext}
where the dot over the letter means  differentiation with respect
to  the variable $\varphi = k x $.

\section{Solution of  Dirac equation in external non-abelian gauge field}

To derive the fermion field $\Psi (x)$ we go from Eq.(1) to the
so-called quadrated Dirac equation which  has the following form:

\begin{widetext}
\begin{eqnarray}
&& \left\{ \partial_\mu \partial^\mu - m^2 + g^2 \left(\gamma^\mu
A^a_\mu T_a \right)^2 - 2 i g \left(\gamma^\mu A^a_\mu T_a \right)
\left(\gamma^\mu \partial_\mu \right) - i g \left(\gamma^\mu k_\mu
 \right) \left(\gamma^\mu
{\dot A}^a_\mu T_a \right) \right\} \Phi (x) = 0 ; \nonumber \\
&& \Psi (x) = \left\{ { i \gamma^{\mu} \left( \partial_{\mu} + i g
\cdot A_{\mu}^a (x) T_a \right) + m \over 2 m} \right\} \Phi (x) .
\end{eqnarray}
\end{widetext}

We will find the solution of the last equation in the form:

\begin{widetext}
\begin{eqnarray}
&&  \Phi (x) \equiv \Phi_{\sigma, \alpha} (x, p) = e^{-ip x} \
\cdot F_{\sigma , \alpha} (k x).
\end{eqnarray}
\end{widetext}
where $F_{\sigma, \alpha} (k x)$ is some multicomponent function
which is a generated Dirac spinor. It depends on  both the  spin
variable $\sigma$ and the variable $\alpha$ which specifies  the
state of a fermion in  the space of the fundamental representation
of the $ SU(N) $ group, thereat  $\alpha = 1 \div  N$.  \ $ p^\nu
= \left( p^0 , {\vec p} \right)$ is the 4-momentum of a particle.

We substitute  $\Phi_{a, \alpha} (x, p)$ in the form  given by
Eq.(9) into the formula (8). Using the relations for the $\gamma$-
matrixes\cite{6} and  Eq.(5), we obtain:

\begin{widetext}
\begin{eqnarray}
&& \left\{ p^2 - m^2 + g^2 {N^2 - 2\over 4N} (  A^a_\mu A^\mu_a )
- 2 g \left( T_a A_\mu^a p^\mu \right)\  - i g  \left(\gamma^\mu
k_\mu
 \right) \left( \gamma^\mu T_a
{\dot A}_\mu^a  \right) \right\} F_{\sigma , \alpha } (\varphi )+
2 i ( p k ) \  {\dot F}_{\sigma , \alpha } (\varphi )  = 0 ;
\nonumber \\
&&  \varphi = k x , \ \ \ \ ( p k) = p^\mu k_\mu ,
\end{eqnarray}
\end{widetext}
where the dot over ${\dot F}_{\sigma , \alpha } (\varphi )$ means
derivative with respect to the variable $\varphi $ as before.

The solution of the derived equation can be formally written as
follows:

\begin{widetext}
\begin{eqnarray}
&& F_{\sigma , \alpha } (\varphi ) = {1\over \sqrt{2 p^0 }} \exp
\left\{ i  {(p^2 - m^2) \varphi  +   g^2 {N^2 - 2\over 4N}
\int\limits_0^\varphi d \varphi^\prime
 ( A_\mu^a ( \varphi^\prime ) A^\mu_a ( \varphi^\prime ) )  \over 2 (p
k) } \right\}\nonumber \\
&& \exp \left\{ - i g \ T_a { \int\limits_0^\varphi d
\varphi^\prime \left( A_\mu^a p^\mu \right)\ + {i\over 2}
\left(\gamma^\mu k_\mu
 \right) \left( \gamma^\mu
 A_\mu^a  \right) \over  ( p k ) } \right\} \ u_\sigma (p) \cdot
 v_\alpha ,
\end{eqnarray}
\end{widetext}
where $u_\sigma (p)$ and $ v_\alpha$  are some spinors which are
the  elements of the  spaces of the corresponding representations.

The last exponent in Eq.(11) is the operator. It is necessary to
determine  acting  such operator on the spinors $u_\sigma (p)$ and
$ v_\alpha$. To do this we expanse the second exponent in Eq.(11)
in the series. After that, we group  the odd and even terms
separately. We also note, that due to Eqs.(6), (7) there are not
any term containing $\gamma$-matrix in the power more than the
first one in the series . The commutative relation (5) as well as
Eqs.(6), (7) allow us to reduce the series which correspond to the
odd and even terms in the expansion of the exponent to the
combinations of sinuses and cosines. As a result we  derive:

\begin{widetext}
\begin{eqnarray}
&&  \exp \left\{ - i g \ T_a { \int\limits_0^\varphi d
\varphi^\prime \left( A_\mu^a p^\mu \right)\ + {i\over 2}
\left(\gamma^\mu k_\mu
 \right) \left( \gamma^\mu
 A_\mu^a  \right) \over  ( p k ) } \right\}  =
 \nonumber \\
&&  \cos \theta \Bigg \{ \left( 1 -  i g  T_a  { \tan \theta \over
\theta ( p k ) }  \  \int\limits_0^\varphi d \varphi^\prime \left(
A_\mu^a p^\mu \right) \right) + {g \left( \gamma^\mu k_\mu \right)
\left( \gamma^\mu A_\mu^a \right) \over 2 ( p k ) } \cdot \Bigg[
{\tan \theta \over \theta}\ T_a  + \nonumber \\
&& {g\over  ( p k) } \ {N^2 - 2\over 4N} \ \int\limits_0^\varphi d
\varphi^\prime \left( A_\mu^a p^\mu \right) \left( - i {\tan
\theta \over \theta} + {g\over  ( p k)}  {\theta - \tan \theta
\over \theta^3} T_b \ \int\limits_0^\varphi d \varphi^\prime
\left( A_\mu^b p^\mu
\right) \right) \Bigg] \Bigg\}; \nonumber \\
&& \theta =  { g \over  ( p k ) } \sqrt{{N^2 - 2\over 4N}} \left(
\int\limits_0^\varphi d \varphi^\prime \left( A_\mu^a (
\varphi^\prime ) \  p^\mu \right) \ \int\limits_0^\varphi d
\varphi^\prime \left( A_a^\mu ( \varphi^{\prime \prime }) \ p_\mu
\right) \right)^{1\over 2}
\end{eqnarray}
\end{widetext}

 Substituting the exponent given by the last formula into Eq.(11) we obtain:

\begin{widetext}
\begin{eqnarray}
&& \Phi_{\sigma, \alpha} (x, p) = e^{-ip x}  F_{\sigma , \alpha }
(\varphi ) = e^{-ip x} \ \cdot {1\over \sqrt{2 p^0 }} \exp \left\{
i  {(p^2 - m^2) \varphi  + g^2 {N^2 - 2\over 4N}
\int\limits_0^\varphi d \varphi^\prime
 ( A_\mu^a ( \varphi^\prime ) A^\mu_a ( \varphi^\prime ) )  \over 2 (p
k) } \right\} \cdot \nonumber \\
&&  \cos \theta \Bigg \{ \left( 1 -  i g  T_a  { \tan \theta \over
\theta ( p k ) }  \  \int\limits_0^\varphi d \varphi^\prime \left(
A_\mu^a p^\mu \right) \right) + {g \left( \gamma^\mu k_\mu \right)
\left( \gamma^\mu A_\mu^a \right) \over 2 ( p k ) } \cdot \Bigg[
{\tan \theta \over \theta}\ T_a  + \nonumber \\
&& {g\over  ( p k) } \ {N^2 - 2\over 4N} \ \int\limits_0^\varphi d
\varphi^\prime \left( A_\mu^a p^\mu \right) \left( - i {\tan
\theta \over \theta} + {g\over  ( p k)}  {\theta - \tan \theta
\over \theta^3} T_b \ \int\limits_0^\varphi d \varphi^\prime
\left( A_\mu^b p^\mu \right) \right) \Bigg] \Bigg\} \ u_\sigma (p)
\cdot
 v_\alpha ,
\end{eqnarray}
\end{widetext}

To determine the spinor $u_\sigma (p)$ we assume that at time
moment $t = - \infty $ there is no interaction of the  fermion
with the external field. Then, $u_\sigma (p)$ is the standard
Dirac spinor satisfying the free Dirac equation. Thereat, we
should note that the fermion is on-shell, so that $p^2 = m^2$.
Since the spinor $u_\sigma (p)$ is  independent on the time
variable it is defined as the spinor of the free Dirac
field\cite{6} at any time moment.

We take the following normalization for $u_\sigma (p)$:

\begin{widetext}
\begin{eqnarray}
&& {\bar u}_\sigma (p) u_\lambda (p^\prime ) = \pm 2m \
\delta_{\sigma  \lambda} \ \delta_{p   p^\prime} ; \ \ \ p^2 =m^2
,
\end{eqnarray}
\end{widetext}
where the plus and minus signs correspond to the Dirac scalar
production of the spinors $u_\sigma (p)$ and $u_\sigma (- p)$,
respectively.

As for the spinor $v_\alpha $ we determine it by the relations:

\begin{widetext}
\begin{eqnarray}
&& v^\dag_\alpha  \ v_\beta = \delta_{\alpha \beta} ; \ \ \ \  Tr
( T_a ) = 0 ; \ \ \ \ \ Tr ( T_a \ T_b ) = {1\over 2} \delta_{a b}
\end{eqnarray}
\end{widetext}

Substituting Eq.(13) into the formula (8) we derive the following
after the direct calculations:

\begin{widetext}
\begin{eqnarray}
&& \Psi (x) =  \Phi_{\sigma, \alpha} (x, p) =  {e^{-ip x}\over
\sqrt{2 p^0 }} \exp \left\{ i  { g^2 {N^2 - 2\over 4N}
\int\limits_0^\varphi d \varphi^\prime
 ( A_\mu^a ( \varphi^\prime ) A^\mu_a ( \varphi^\prime ) )  \over 2 (p
k) } \right\}\nonumber \\
&&  \cos \theta \Bigg \{ \left( 1 -  i g  T_a  { \tan \theta \over
\theta ( p k ) }  \  \int\limits_0^\varphi d \varphi^\prime \left(
A_\mu^a p^\mu \right) \right) + {g \left( \gamma^\mu k_\mu \right)
\left( \gamma^\mu A_\mu^a \right) \over 2 ( p k ) } \cdot \Bigg[
{\tan \theta \over \theta}\ T_a  + \nonumber \\
&& {g\over  ( p k) } \ {N^2 - 2\over 4N} \ \int\limits_0^\varphi d
\varphi^\prime \left( A_\mu^a p^\mu \right) \left( - i {\tan
\theta \over \theta} + {g\over  ( p k)}  {\theta - \tan \theta
\over \theta^3} T_b \ \int\limits_0^\varphi d \varphi^\prime
\left( A_\mu^b p^\mu \right) \right) \Bigg] \Bigg\}\ u_\sigma (p)
\cdot
 v_\alpha ;  \nonumber \\
 && \theta =  { g \over  ( p k ) } \sqrt{{N^2 - 2\over 4N}} \left(
\int\limits_0^\varphi d \varphi^\prime \left( A_\mu^a (
\varphi^\prime ) \  p^\mu \right) \ \int\limits_0^\varphi d
\varphi^\prime \left( A_a^\mu ( \varphi^{\prime \prime }) \ p_\mu
\right) \right)^{1\over 2} ; \ \ \ \  p^2 = p^\mu p_\mu = m^2 .
\end{eqnarray}
\end{widetext}

The derived $\Psi (x)$-function  is the exact solution of the
Dirac equation in the external non-abelian field having the form
of a plane wave on the light cone. The function (16) is normalized
by the $\delta$-function as follows:

\begin{widetext}
\begin{eqnarray}
&& \int d^3 x  \Psi^\dag (x, p^\prime ) \Psi (x, p)   = (2\pi )^3
\delta^3 ( {\vec p} - {\vec p^\prime }  ).
\end{eqnarray}
\end{widetext}

 The direct calculation
shows that $\Phi_{\sigma, \alpha} (x, p)$ and $\Phi_{- \sigma,
\alpha} (x, - p)$ are orthogonal. In this way, it is obvious, that
$\Phi_{\sigma, \alpha} (x, p)$ is the so-called  positive
frequency function whereas $\Phi_{\sigma, \alpha} (x, p)$ is
negative frequency one\cite{6,7}. This fact allows us to construct
the general solution of the Dirac equation which describes  the
states both particles and anti-particles. Combining the functions
$\Phi_{\sigma, \alpha} (x, p)$ and $\Phi_{- \sigma, \alpha} (x, -
p)$, we obtain:

\begin{widetext}
\begin{eqnarray}
&& \Psi (x) = \sum\limits_{\sigma ,  \alpha } \int {d^3 p \over
\sqrt{2p^0 \ } (2\pi)^3 } \left\{ {\hat a}_{\sigma , \alpha} (p)
\Psi_{\sigma , \alpha }  ( x, p ) +
{\hat b}^\dag_{\sigma , \alpha } (p) \Psi_{ - \sigma , \alpha }  ( x, - p )  \right\} \nonumber \\
&& {\bar \Psi (x)}  = \sum\limits_{\sigma ,  \alpha} \int {d^3 p
\over \sqrt{2p^0 \ } (2\pi)^3 } \left\{ {\hat a}^\dag_{\sigma ,
\alpha} (p)\ {\bar \Psi}_{\sigma , \alpha }  ( x, p ) + {\hat
b}_{\sigma , \alpha} (p) \ {\bar \Psi}_{ - \sigma , \alpha }  ( x,
 - p ) \right\} ,
\end{eqnarray}
\end{widetext}
where the symbols ${\hat a}^\dag_{\sigma , \alpha} (p) ;{\hat
b}^\dag_{\sigma , \alpha} (p)$ and ${\hat a}_{\sigma , \alpha}
(p); {\hat b}_{\sigma , \alpha} (p)$ are the operators of creation
and cancellation of a fermion (${\hat a}_{\sigma , \alpha} (p)
;{\hat a}^\dag_{\sigma , \alpha} (p)$) and anti-fermion (${\hat
b}_{\sigma , \alpha} (p) ;{\hat b}^\dag_{\sigma , \alpha} (p)$) ,
respectively\cite{6,7}. Thereat,  ${\hat a}_{\sigma , \alpha} (p)$
¨ $ {\hat a}^\dag_{\sigma , \alpha} (p)$;    ${\hat b}_{\sigma ,
\alpha} (p)$ ¨ $ {\hat b}^\dag_{\sigma , \alpha} (p)$ satisfy the
standard commutative relations for fermion operators.

\section{Conclusion }

A relativistic fermion with the spin $1/2$  in a external $SU(N)$
gauge field is considered. The exact solution of the Dirac
equation where the  field is in the form of a plane wave on the
light cone is derived. The whole set of the solution which
corresponds to both particles and anti-particles is obtained.

\end{document}